\documentclass{llncs}
\RequirePackage{filecontents}
\RequirePackage{comgipp}

\newcommand{\theReferenceText}{\fullcite{GreinerPetter2019}}
\RequirePackage{amsmath}

\usepackage[T1]{fontenc}
\usepackage{booktabs} %

\usepackage[utf8]{inputenc}
\usepackage{hyperref}

\usepackage{graphicx}
\graphicspath{{images/}}

\usepackage{wrapfig}
\usepackage{array}

\usepackage{color}

\definecolor{gray}{rgb}{0.4,0.4,0.4}
\definecolor{darkblue}{rgb}{0.0,0.0,0.6}
\definecolor{cyan}{rgb}{0.0,0.6,0.6}
\definecolor{darkgreen}{rgb}{0,0.5,0}

\usepackage{listings}
\usepackage{graphics}

\lstdefinelanguage{XML}
{
  morecomment=[f][\color{red}]{-\ },         %
  morestring=[b]",
  morestring=[s]{>}{<},
  morecomment=[s]{<?}{?>},
  morecomment=[s]{!--}{--},
  morecomment=[s][\color{red}]{\$}{\$},
  commentstyle=\color{darkgreen},
  stringstyle=\color{black},
  identifierstyle=\color{darkblue},
  keywordstyle=\color{cyan},
  morecomment=[f][\color{green}]+,       %
  morekeywords={xmlns,version,type}%
}

\usepackage[
 backend=biber,
 style=numeric,
 url=false,
 isbn=false,
 doi=false,
 bibencoding=utf8,
 maxbibnames=2,
 giveninits=true
]{biblatex}
\usepackage[
  citations=true,
  hybrid=true
   ]{markdown}
\markdownSetup{
  renderers = {
    link = {\href{#2}{#1}}
  }
}

\begin{document}
\mainmatter

\title{Why Machines Cannot Learn Mathematics, Yet}
\titlerunning{Why Machines Cannot Learn Mathematics}

\author{%
    Andr\'{e} Greiner-Petter\inst{1}, %
    Terry Ruas\inst{2}, %
    Moritz Schubotz\inst{1},\\
    Akiko Aizawa\inst{3}, %
    William Grosky\inst{2}, %
    Bela Gipp\inst{1}%
}

\authorrunning{A. Greiner-Petter, T. Ruas, M. Schubotz, et al.}

\institute{%
	University of Wuppertal, Wuppertal, Germany\\
	\email{\{last\}@uni-wuppertal.de}
	\and University of Michigan-Dearborn, Dearborn, USA\\
	\email{\{truas,wgrosky\}@umich.edu}
	\and National Institute of Informatics, Tokyo, Japan\\
	\email{aizawa@nii.ac.jp}
}

\maketitle
\ifdefined\theReferenceText
\thispagestyle{firststyle}
\fi
\begin{abstract}
Nowadays, Machine Learning (ML) is seen as the universal solution to improve the effectiveness of information retrieval (IR) methods. However, while mathematics is a precise and accurate science, it is usually expressed by less accurate and imprecise descriptions, contributing to the relative dearth of machine learning applications for IR in this domain. Generally, mathematical documents communicate their knowledge with an ambiguous, context-dependent, and non-formal language. Given recent advances in ML, it seems canonical to apply ML techniques to represent and retrieve mathematics semantically. In this work, we apply popular text embedding techniques to the arXiv collection of STEM documents and explore how these are unable to properly understand mathematics from that corpus. In addition, we also investigate the missing aspects that would allow mathematics to be learned by computers.\relax
\end{abstract}

\keywords{Mathematical Information Retrieval, Machine Learning,
Word Embeddings, Math Embeddings, Mathematical Objects of Interest}

\begin{figure}[ht] \centering \includegraphics[width=\columnwidth]{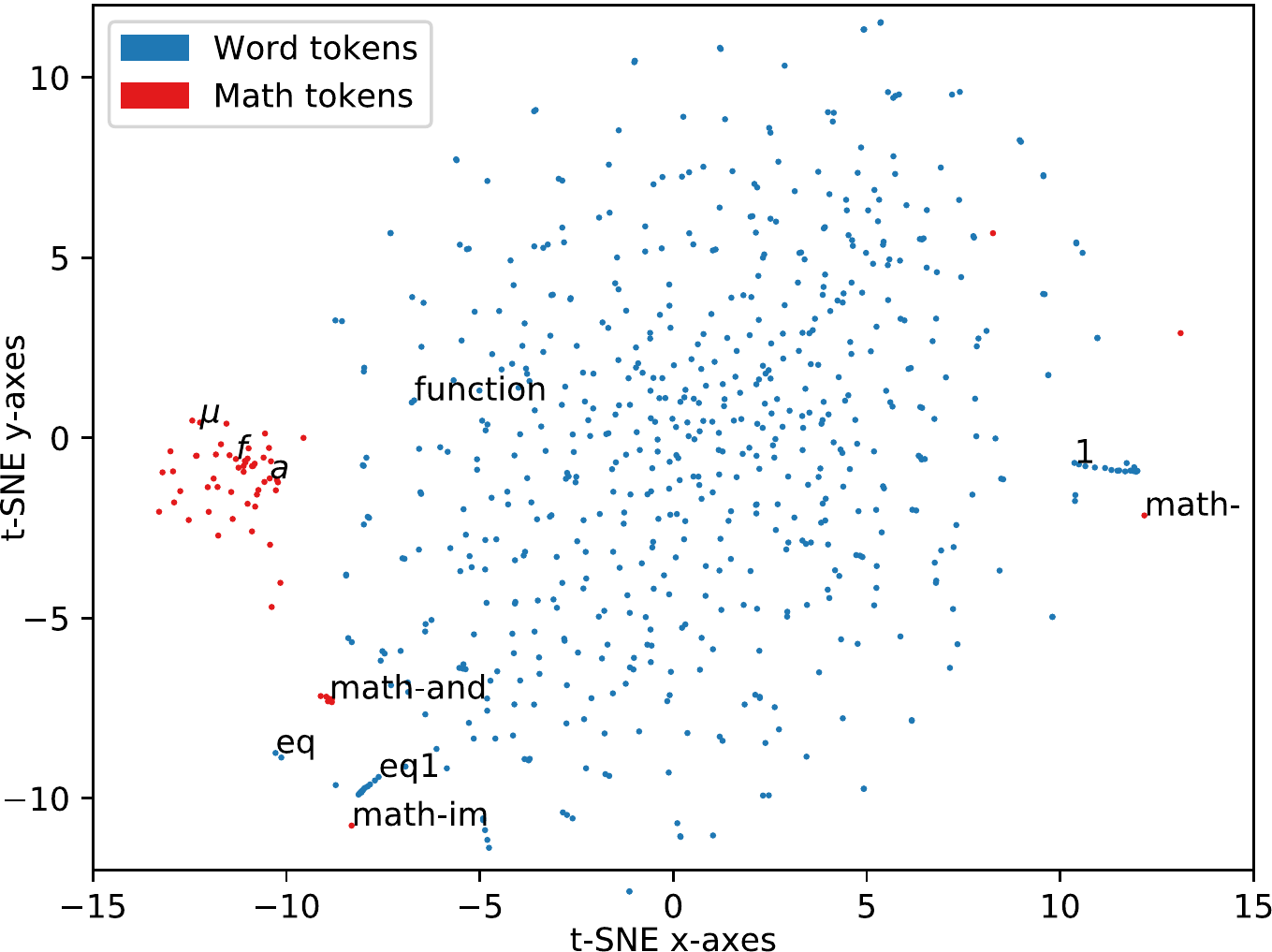} \caption{t-SNE plot of top 1000 closest vectors of the identifier $f$. For this plot, we used a perplexity of 80 and the default values of the TSNE python package for other settings.} \label{fig:tsne} \end{figure}

\markdownRendererHeadingTwo{Introduction}\markdownRendererInterblockSeparator
{}Mathematics is capable of explaining complex concepts and relations in a compact, precise, and accurate way. Learning this idiom takes time and is often difficult, even to humans. The general applicability of mathematics allows a certain level of ambiguity in its expressions. This ambiguity is regularly mitigated by short explanations following or preceding these mathematical expressions, that serve as context to the reader. Along with context dependency, inherent issues of linguistics (e.g. ambiguity, non-formality) make it even more challenging for computers to understand mathematical expressions. Said that, a system capable of capturing the semantics of mathematical expressions automatically would be suitable for several applications, from improving search engines to recommender systems. \markdownRendererInterblockSeparator
{}During our evaluations of MathMLBen~\cite{SchubotzGSMCG18}, a benchmark for converting mathematical \LaTeX{} expressions into MathML, is possible to notice several fundamental problems that generally affect prominent ML approaches to learn semantics of mathematical expressions. For instance, the first entry of the benchmark,\markdownRendererInterblockSeparator
{}\begin{align}\label{eq:waerden} W(2, k) > 2^k/k^\varepsilon \end{align}\markdownRendererInterblockSeparator
{}\noindent is extracted from the English Wikipedia page about Van der Waerden's theorem\footnote{\url{https://en.wikipedia.org/wiki/Van_der_Waerden's_theorem}}. Without further explanation, the symbols $W$, $k$, and $\varepsilon$ might have several possible meanings. Depending on which one is considered, even the structure of the formula may be different. If we consider $W$ as a variable, instead of a function, it changes the interpretation of $W(2,k)$ to a multiplication operation. \markdownRendererInterblockSeparator
{}Learning connections, such as between $W$ and the entity `\markdownRendererEmphasis{Van der Waerden's number}', requires a large specifically labeled scientific database that contains these mathematical objects. Furthermore, a fundamental understanding of the mathematical expression would increase the performance during the learning process, e.g., that $W(2,k)$ and $W(n,k)$ contain the same function.\markdownRendererInterblockSeparator
{}Word embedding techniques has received significant attention over the last years in the Natural Language Processing (NLP) community, especially after the publication of word2vec~\cite{Mikolov-b:13}. Recently, more and more projects try to adapt this knowledge for solving Mathematical Information Retrieval (MIR) tasks~\cite{DBLP:journals/corr/GaoJYYYT17,DBLP:journals/corr/abs-1803-09123,DBLP:conf/mkm/YoussefM18,DBLP:journals/corr/abs-1902-06034}. While all of these projects follow similar approaches and obtain promising results, all of them fail to understand mathematical expressions because of the same fundamental issues. In this paper, we explore some of the main aspects that we believe are necessary to leverage the learning of mathematics by computer systems. We explain, with our evaluations of word embedding techniques on the arXMLiv 2018~\cite{SML:arXMLiv:08.2018} dataset, why current ML approaches are not applicable for MIR tasks, yet.\markdownRendererInterblockSeparator
{}\markdownRendererHeadingTwo{Background \& Related Work \label{sec:related-work}}\markdownRendererInterblockSeparator
{}Understanding mathematical expressions essentially means comprehending the semantic value of its internal components, which can be accomplished by linking its elements with their corresponding mathematical definitions. Current MIR approaches~\cite{kristianto2014extracting,disSigir16,Schubotz2017} try to extract textual descriptors of the parts that compose mathematical equations. Intuitively, there are questions that arise from this scenario, such as (i) how to determine the parts which have their own descriptors, and (ii) how to identify correct descriptors over others. \markdownRendererInterblockSeparator
{}Answers to (i) are more concerned in choosing the correct definitions for which parts of a mathematical expression should be considered as one mathematical object~\cite{DBLP:conf/lwa/Kohlhase17,POM-Tagger,SchubotzGSMCG18}. Current definitions, such as the content MathML 3.0\footnote{\url{https://www.w3.org/TR/MathML3/}} specification, are often imprecise\footnote{Note that OpenMath is another specification specifically designed for encoding semantics of mathematics. However, content MathML is an encoding of OpenMath and inherent problems of content MathML also apply for OpenMath (see \url{https://www.openmath.org/om-mml/}).}. For example, content MathML 3.0 uses \verb|csymbol| elements for functions and specifies them as expressions that \markdownRendererEmphasis{refer to a specific, mathematically-defined concept with an external definition}\footnote{\url{https://www.w3.org/TR/MathML3/chapter4.html##contm.csymbol}}. However, it is not clear whether $W$ or the sequence $W(2,k)$ (Equation~\ref{eq:waerden}) should be declared as a \verb|csymbol|. Another example are content identifiers, which MathML specifies as \markdownRendererEmphasis{mathematical variables which have properties, but no fixed value}\footnote{\url{https://www.w3.org/TR/MathML3/chapter4.html##contm.ci}}. While content identifiers are allowed to have complex rendered structures (e.g.~$\beta^2_{i}$), it is not permitted to enclose identifiers within other identifiers. Let us consider $\alpha_{i}$, where $\alpha$ is a vector and $\alpha_{i}$ its $i$-th element. In this case, $\alpha_{i}$ should be considered as a composition of three content identifiers, each one carrying its own individualized semantic information, namely the vector $\alpha$, the element $\alpha_{i}$ of the vector, and the index $i$. However, with the current specification, the definition of these identifiers would not be canonical. One possible workaround to represent such expressions with content MathML is to use a structure of four nodes, interpreting $\alpha_{i}$ as a function with the \verb|csymbol| \markdownRendererEmphasis{vector-selector}. However, ML algorithms and MIR approaches would benefit from more precise definitions and a unified answer for (i). Most of the related work relies on these relatively vague definitions and in the analysis of content identifiers, focusing their efforts on (ii). \markdownRendererInterblockSeparator
{}In~\cite{disSigir16}, an approach is presented for scoring pairs of identifiers and \textit{definiens}\footnote{\textit{definiens} is a phrase that defines an identifier or mathematical object. Considering equation~\eqref{eq:waerden}, the correct definiens for $W$ is the phrase '\markdownRendererEmphasis{Van der Waerden's number}'.} by the number of words between them. Their approach is based on the assumption that correct definiens appear close to the identifier and to the complex mathematical expression that contains this same identifier. Kristianto et al.~\cite{kristianto2014extracting} introduce an ML approach, in which they train a Support Vector Machine (SVM) to consider sentence patterns and other characteristics as features (e.g. part-of-speech (POS) tags, parse trees). Later,~\cite{Schubotz2017} combine the aforementioned approaches and use pattern recognition based on the POS tags of common identifier-definiens pairs, the distance measurements, and SVM, reporting results for precision and recall of 48.60\% and 28.06\%, respectively. These results can be considered as a baseline for MIR tasks.\markdownRendererInterblockSeparator
{}More recently, some projects try to use embedding techniques to learn patterns of the correlations between context and mathematics. In the work of ~\cite{DBLP:journals/corr/GaoJYYYT17}, they embed single symbols and train a model that is able to discover similarities between mathematical symbols. Similarly to this approach, Krstovski and Blei \cite{DBLP:journals/corr/abs-1803-09123} use a variation of word embeddings (briefly discussed in Section~\ref{sec:embed}) to represent complex mathematical expressions as single unit tokens for IR. In 2019, M. Yasunaga et al.~\cite{DBLP:journals/corr/abs-1902-06034} explore an embedding technique based on recurrent neural networks to improve topic models by considering mathematical expressions. They state their approach outperforms topic models that do not consider mathematics in text and report a topic coherence improvement of $0.012$ over the LDA\footnote{Latent Dirichlet Allocation} baseline. What all these embedding projects have in common is that they show promising examples and suppose a high potential, but do not evaluate their results for MIR. \markdownRendererInterblockSeparator
{}Questions (i), (ii), and other pragmatic issues are already in discussion in a bigger context, as data production continues to rise and digital repositories seem to be the future for any archive structure. The National Research Council is making efforts to establish what they call the \markdownRendererEmphasis{Digital Mathematics Library} (DML)\footnote{\url{https://www.nap.edu/read/18619}}, a project under the International Mathematical Union. The goal of this future project is to take advantage of new technologies and help to solve the inability to search, relate, and aggregate information about mathematical expressions in documents over the web. \markdownRendererInterblockSeparator
{}\markdownRendererHeadingTwo{Machine Learning on Embeddings \label{sec:embed}}\markdownRendererInterblockSeparator
{}The \emph{word2vec}~\cite{Mikolov-b:13} technique computes real-valued vectors for words in a document using two main approaches: skip-gram and continuous bag-of-words (CBOW). Both produce a fixed length $n$-dimensional vector representation for each word in a corpus. In the skip-gram training model, one tries to predict the context of a given word, while CBOW predicts a target word given its context. In word2vec, context is defined as the adjacent neighboring words in a defined range, called a sliding window. The main idea is that the numerical vectors representing similar words should have close values if the words have similar context, often illustrated by the \markdownRendererEmphasis{king-queen} relationship\footnote{$\vec{v}_{\text{king}}-\vec{v}_{\text{man}} \approx \vec{v}_{\text{queen}}-\vec{v}_{\text{woman}}$}.\markdownRendererInterblockSeparator
{}Extending word2vec's approaches, Le and Mikolov~\cite{Le:14} propose~\emph{Paragraph Vectors} (PV), a framework that learns continuous distributed vector representations for any size of text segment (e.g. sentences, paragraphs, documents). This technique alleviates the inability of word2vec to embed documents as one single entity. This technique also comes in two distinct variations: Distributed Memory (DM) and Distributed Bag-of-Words (DBOW), which are analogous to the skip-gram and CBOW training models respectively. However, in both approaches, an extra feature vector representing the text segment, named paragraph-id, is included as another word. This paragraph-id is updated throughout the entire document, based on the current evaluated context window for each word, and is used to represent the whole text segment.\markdownRendererInterblockSeparator
{}Recently, researchers have been trying to improve their semantic representations, producing multiple vectors (multi-sense embeddings) based on the word's sense, context, and distribution in the corpus~\cite{Huang:12,Reisinger:10}. Another concern with traditional techniques is that they often neglect exploring lexical structures with valuable prior knowledge about the semantic relations, such as: WordNet~\cite{DBLP:journals/cacm/Miller95}, ConceptNet~\cite{LiuCN:04} and BabelNet~\cite{Navigli:12}. These lexical structures offer a rich semantic environment that illustrate the word-senses, their use, and how they relate to each other. Some publications take advantage of the robustness provided by word embeddings approaches and lexical structures to combine them into multi-sense representations, improving their overall performance in many NLP downstream tasks~\cite{Mancini:17,Ruas:19,Taher:16}.\markdownRendererInterblockSeparator
{}The lack of solid references and applications that provide the same semantic structure of natural language for mathematical identifiers make their disambiguation process even more challenging. In natural texts, one can try to infer the most suitable word sense for a word based on the lemma\footnote{canonical form, dictionary form, or citation form of a set of words} itself, the adjacent words, dictionaries, thesaurus and so on. However, in the mathematical arena, the scarcity of resources and the flexibility of redefining their identifiers take this issue to a more delicate scenario. The context text preceding or following the mathematical equation is essential for its understanding.\markdownRendererInterblockSeparator
{}More recently,~\cite{DBLP:journals/corr/abs-1803-09123} propose a variation of word embeddings for mathematical expressions. Their main idea relies on the construction of a distributed representation of equations, considering the word context vector of an observed word and its word-equation context window. They treat equations as single-unit words (EqEmb), which eventually appears in the context of different words. They also try to explore the effects of considering the elements of mathematical expressions separately (EqEmb-U). In this scenario, mathematical equations are represented using a Syntax Layout Tree (SLT)~\cite{DBLP:conf/sigir/ZanibbiDKT16}, which contains the spatial relationship between its symbols. While they present some interesting findings for retrieving entire equations, little is said about the vectors representing equation units and how they are described in their model. The word embedding techniques seem to have potential for semantic distance measures between complex mathematical expressions. However, they are not appropriate for extracting semantics of identifiers separately. This is an indication that the problems of representing mathematical identifiers are tied to more fundamental issues, which are explained in Section~\ref{sec:learnable}.\markdownRendererInterblockSeparator
{}Since the overall performance of word embedding algorithms has shown superior results in many different NLP tasks, such as machine translation~\cite{Mikolov-b:13}, relation similarity~\cite{Iacobacci:15}, word sense disambiguation~\cite{Camachob:15}, word similarity~\cite{Neela:14,Ruas:19}, and topic categorization~\cite{Taher:17}. In the same direction, we also explore how well mathematical tokens can be embedded according to their semantic information. However, mathematical formulae are highly ambiguous and if not properly processed, their representation is jeopardized.\markdownRendererInterblockSeparator
{}\markdownRendererHeadingThree{How to Embed Mathematics \label{subsec:math-embed}}\markdownRendererInterblockSeparator
{}There are two main standard formats in which to represent mathematics in science: \LaTeX{} and MathML. The former is used by humans for writing scientific documents. The latter, on the other hand, is popular in web representations of mathematics due to its machine readability and XML structure. There has been a major effort to automatically convert \LaTeX{} expressions to MathML~\cite{SchubotzGSMCG18} ones. However, neither \LaTeX{} nor MathML are practical formats for embeddings. Considering the equation embedding techniques in~\cite{DBLP:journals/corr/abs-1803-09123}, we devise three main types of mathematical embeddings.\markdownRendererInterblockSeparator
{}\noindent \markdownRendererStrongEmphasis{Mathematical Expressions as Single Tokens:} EqEmb~\cite{DBLP:journals/corr/abs-1803-09123} uses entire mathematical expressions as one token. In this type, the inner structure of the mathematical expression is not taken into account. For example, Equation~\eqref{eq:waerden} is represented as one single token $t_1$. Any other expression, such as $W(2,k)$ in the surrounding text of~\eqref{eq:waerden}, is an entirely independent token $t_2$. Therefore, this approach does not learn any connections between $W(2,k)$ and \eqref{eq:waerden}. While this approach seems to hold interesting results for comparing mathematical expressions, it fails in representing the semantic aspects of inner elements in mathematical equations. \markdownRendererInterblockSeparator
{}\noindent \markdownRendererStrongEmphasis{Stream of Tokens:\label{subsec:identifiertokens}} Instead of embedding mathematical expressions as a single token, we can represent them through a sequence of its inner tokens. For example, considering only the identifiers in Equation~\eqref{eq:waerden}, we would have a stream of three tokens $W$, $k$, and $\varepsilon$. This approach has the advantage of learning all mathematical tokens. However, this method also has some drawbacks. Complex mathematical expressions may lead to long chains of elements, which can be especially problematic when the window size of the training model is too small. Naturally, there are approaches to reduce the length of chains. In Section~\ref{sec:learnable} we show our own model which uses a stream of mathematical identifiers and cut out all other expressions. In~\cite{DBLP:journals/corr/GaoJYYYT17}, L.~Gao et al. use a CBOW and embed all mathematical symbols, including identifiers and operands, such as $+$, $-$ or variations of equalities $=$. In~\cite{DBLP:journals/corr/abs-1902-06034}, they do not cut out symbols and train their model on the entire sequence of tokens that the \LaTeX{} tokenizer generates. Considering Equation~\eqref{eq:waerden}, it would result in a stream of 13 tokens. They use a long short-term memory (LSTM) architecture to handle longer chains of tokens and also to limit their length to $20-150$ tokens. Usually, in word embeddings, such behaviour is not preferred since it increases the noise in the data\footnote{Noise means, the data consists of many uninteresting tokens that affect the trained model negatively.}. We will see later in the paper (Section~\ref{subsec:math-embed}), that a typically trained model on mathematical embeddings is able to detect similarities between mathematical objects but do not perform well detecting connections to word descriptors. Therefore, we consider close relations of mathematical symbols to other mathematical symbols as noise. To mitigate this issue, we only work with mathematical identifiers and not any other symbols or structures.\markdownRendererInterblockSeparator
{}\noindent \markdownRendererStrongEmphasis{Semantic Groups of Tokens:} The third approach of embedding mathematics is only theoretical, and concerns the aforementioned problems related to the vague definitions of identifiers and functions in a standardized format (e.g. MathML). As previously discussed, current MIR and ML approaches would benefit from a basic structural knowledge of mathematical expressions, such that variations of function calls (e.g. $W(r,k)$ and $W(2,k)$) can be recognized as the same function. Instead of defining a unified standard, current techniques use their own ad-hoc interpretations of structural connections, e.g., $\alpha_i$ is one identifier rather than three~\cite{SchubotzGSMCG18,Schubotz2017}. We assume that an embedding technique would benefit from a system that is able to detect the parts of interest in mathematical expressions prior any training processes. However, such system still does not exist.\markdownRendererInterblockSeparator
{}\markdownRendererHeadingTwo{Performance of Math Embeddings \label{subsec:performance}}\markdownRendererInterblockSeparator
{}The examples illustrated in~\cite{DBLP:journals/corr/GaoJYYYT17,DBLP:journals/corr/abs-1803-09123,DBLP:journals/corr/abs-1902-06034} seem to be feasible as a new approach for distance calculations between complex mathematical expressions. While comparing mathematical expressions is essentially practical for search engines or automatic plagiarism detection systems, these approaches do not seem to capture the components of complex structure separately, which are necessary for other applications, such as automated reasoning. Another aspect to be considered is that in~\cite{DBLP:journals/corr/abs-1803-09123} they do not train mathematical identifiers, preventing their system from learning connections between identifiers and definiens (e.g., $W(2,k)$ and the definiens `\markdownRendererEmphasis{Van der Waerden number}'). Additionally, the connection between entire equations and definiens is, at some level, questionable. Entire equations are rarely explicitly named\footnote{However, it is common for groundbreaking findings, such as \markdownRendererEmphasis{Pythagorean's theorem} or the \markdownRendererEmphasis{energy-mass equivalence}.}. However, in the extension EqEmb-U~\cite{DBLP:journals/corr/abs-1803-09123}, they use an SLT representation to tokenize mathematical equations and to obtain specific unit-vectors, which is similar to our \markdownRendererEmphasis{identifiers as tokens} approach.\markdownRendererInterblockSeparator
{}In order to investigate the discussed approaches, we apply variations of a word2vec implementation to extract mathematical relations from the arXMLiv 2018~\cite{SML:arXMLiv:08.2018} dataset, an HTML collection of the arXiv.org preprint archive\footnote{\url{https://arxiv.org/}}, which is used as our training corpus. We also consider the subsets that do not report errors during the document conversion (i.e. \markdownRendererEmphasis{no\textunderscore problem} and \markdownRendererEmphasis{warning}) which represent 70\% of archive.org. There are other approaches that also produce word embeddings given a training corpus as an input, such as fastText~\cite{Bojanowski:17}, ELMo~\cite{Matthew:18}, and GloVe \cite{Penni:14}. The choice for word2vec is justified because of its implementation, general applicability, and robustness in several NLP tasks~\cite{Iacobacci:15,Iacobacci:16,Li:15,Mancini:17,Taher:16,Ruas:19}. Additionally, in fastText they propose to learn word representations as a sum of the $n$-grams of its constituent characters (sub-words). This would incorporate a certain noise to our experiments. In ELMo, they compute their word vectors as the average of their characters representations, which are obtained through a two-layer bidirectional language model (biLM). This would bring even more granularity than fastText, as they consider each character in a word as having their own $n$-dimensional vector representation. Another factor that prevent us from using ELMo, for now, is its expensive training process\footnote{\url{https://github.com/allenai/bilm-tf}}. Closer to the word2vec technique, GloVe~\cite{Penni:14} is also considered, but its co-occurrence matrix would escalate the memory usage, making its training for arXiv not possible at the moment. We also examine the recently published Universal Sentence Encoder (USE)~\cite{Cer:18} from google, but their implementation does not allow one to use a new training corpus, only to access its pre-calculated vectors.\markdownRendererInterblockSeparator
{}As a pre-processing step, mathematical expressions are represented using MathML\footnote{The source \TeX{} file has to use mathematical environments for its expressions.} notation. Firstly, we replace all mathematical expressions by the sequence of the identifiers it contains, i.e., $W(2,k)$ is replaced by `$W$ $k$'. Secondly, we remove all common English stopwords from the training corpus. Finally, we train a word2vec model (skip-gram) using the following hyperparameters configuration\footnote{Non mentioned hyperparameters are used with their default values as described in the Gensim API~\cite{rehurek_lrec}}: vector size of 300 dimensions, a window size of 15, minimum word count of 10, and a negative sampling of $1E-5$.\markdownRendererInterblockSeparator
{}The trained model is able to partially incorporate semantics of mathematical identifiers. For instance, the closest\footnote{Considering cosine similarity.} 27 vectors to the mathematical identifier $f$ are mathematical identifiers themselves and the fourth closest noun vector to $f$ is $\vec{v}_{\text{function}}$. Inspired by the classic \markdownRendererEmphasis{king-queen} example, we explore which tokens perform best to model a known relation. Consider an approximation $\vec{v}_{\text{variable}}-\vec{v}_{\text{a}} \approx \vec{v}-\vec{v}_{\text{f}}$, where $\vec{v}_{\text{variable}}$ represents the word \markdownRendererEmphasis{variable}, $\vec{v}_{\text{a}}$ the identifier $a$, and $\vec{v}_{\text{f}}$ represents $f$. We are looking for $\vec{v}$ that fits best for the approximation. We call this measure the \markdownRendererEmphasis{semantic distance} to $f$ with respect to a given relation between two vectors. Table~\ref{tab:semantic-dist} shows the top 10 semantically closest results to $f$ with respect to the relation between $\vec{v}_{\text{a}}$ and $\vec{v}_{\text{variable}}$.\markdownRendererInterblockSeparator
{}\setlength\intextsep{0.5cm} \setlength\columnsep{0.7cm} \begin{wraptable}{r}{0pt} \begin{tabular}{|c|c|} \hline \textbf{Tokens} & \textbf{Cosine Distances} \\\hline variables & $0.7600$ \\\hline function & $0.7154$ \\\hline appropriate & $0.6925$ \\\hline independent & $0.6789$ \\\hline instead & $0.6784$ \\\hline defined & $0.6729$ \\\hline namely & $0.6719$ \\\hline continuous & $0.6707$ \\\hline depends & $0.6629$ \\\hline represents & $0.6623$ \\\hline \end{tabular} \caption{Semantically closest 10 results to $f$ with respect to the relation between $\vec{v}_{\text{a}}$ and $\vec{v}_{\text{variable}}$.} \label{tab:semantic-dist} \end{wraptable}\markdownRendererInterblockSeparator
{}We also perform an extensive evaluation on the first 100 entries\footnote{Same entries used in~\cite{Schubotz2017}} of the \markdownRendererEmphasis{MathMLBen} benchmark~\cite{SchubotzGSMCG18}. We evaluate the average of the \markdownRendererEmphasis{semantic distances} with respect to the relations between $\vec{v}_{\text{variable}}$ and $\vec{v}_{\text{x}}$, $\vec{v}_{\text{variable}}$ and $\vec{v}_{\text{a}}$, and $\vec{v}_{\text{function}}$ and $\vec{v}_{\text{f}}$. In addition, we consider only results with a cosine similarity of $0.7$ or above to maintain a minimum quality in our results. The overall results were poor with a precision of $p = .0023$ and a recall of $r = .052$. For the identifier $W$ (Equation~\eqref{eq:waerden}), the evaluation presents four semantically close results: \markdownRendererEmphasis{functions}, \markdownRendererEmphasis{variables}, \markdownRendererEmphasis{form}, and the mathematical identifier $q$. Even though expected, the scale of the presented results are astonishing. \markdownRendererInterblockSeparator
{}Additionally, we also try the Distributed Bag-of-Words of Paragraph Vectors (DBOW-PV)~\cite{Le:14} considering the approach of~\cite{Schubotz2017}. In~\cite{Schubotz2017}, they analyze all occurrences of mathematical identifiers and consider the entire article at once. We assume this prevents the algorithm from finding the right descriptor in the text, since later or prior occurrences of an identifier might appear in a different context, and therefore potentially introduce different meanings. Instead of using the entire document, we consider the algorithm of~\cite{Schubotz2017} only in the input paragraph and similar paragraphs given by our DBOW-PV model. Unfortunately, the obtained variance within the paragraphs brings a high number of false positives to the list of candidates, which affects negatively our performance.\markdownRendererInterblockSeparator
{}We also experiment other hyperparameters when training our word embeddings model to see if it is possible to improve the overall results. However, while the performance decreases, no drastic structural changes appear in the model. Figure~\ref{fig:tsne} illustrates a t-SNE plot\footnote{Note that t-SNE plots may misleadingly create clusters that do not exists in the model. To overcome this issue we create several plots with different settings. The results remain similar to the plot we show which is an indication that the visual clusters exists also in the model.} of the model trained with 400 dimensions, a window size of 25, and minimum count of 10 words, without any filters applied to the text. The plot is similar to the visualized model presented in~\cite{DBLP:journals/corr/GaoJYYYT17}, even though they use a different embedding technique. Compared to~\cite{DBLP:journals/corr/GaoJYYYT17}, we provide a bigger picture of the model that reveals some dense clusters for numbers at $(11,-1)$ with the math token for \markdownRendererEmphasis{invisible times} nearby, equation abbreviations such as, \textit{eq1}, at $(-8,-9)$, and logical operators at $(-9,-7)$. We highlight mathematical tokens in the model in red and word tokens in blue. The plot in Figure~\ref{fig:tsne} illustrates that mathematical tokens are close to each other.\markdownRendererInterblockSeparator
{}\markdownRendererInterblockSeparator
{}Based on the presented results, one can still argue that more settings should be explored (e.g. different embedding techniques, parameters) for the embeddings phase and different pre-processing steps (e.g. stemming and lemmatization) be adopted. This would probably solve some minor problems, such as removing the inaccurate first hit in Table~\ref{tab:semantic-dist}. Nevertheless, the overall results would not be improved to a point of being comparable to~\cite{Schubotz2017} findings, which report a precision of $p=0.48$. The main reason for this is that, mathematics as a language is highly customizable. Many of the defined relations between mathematical concepts and their descriptors are only valid in a local scope. Consider, for example, an author that notes his algorithm by $\pi$. This does not change the general meaning of $\pi$, even though it effects the meaning in the scope of the article. Current ML approaches only learn patterns of most frequently used combinations, e.g., between $f$ and \textit{function}, as seen in Table~\ref{tab:semantic-dist}. Furthermore, we assume this is a general problem that different embedding techniques and tweaks of settings, such as those illustrated in Section~\ref{sec:related-work}, would not solve. Therefore, in the following section, we present some concepts that we believe can help ML algorithms to better understand mathematics.\markdownRendererInterblockSeparator
{}\markdownRendererHeadingTwo{Make Math Machine Learnable \label{sec:learnable}}\markdownRendererInterblockSeparator
{}A case study~\cite{WolskaDML10} has shown that 70\% of mathematical symbols are explicitly declared in the context. Only four reasons are causing an explicit declaration in the context: (a) a new mathematical symbol is defined, (b) a known notation is changed, (c) used symbols are present in other contexts and require specifications to be properly interpreted, or (d) authors declarations were redundant (e.g. for improving readability). We assume (d) is a rare scenario compared to (a-c), unless in educational literature. Current math-embedding techniques can learn semantic connections only in those 70\%, where the definiens are available. Besides (d), the algorithm would learn either rare notations (in case of (a)) or ambiguous notations (in cases (b-c)). The flexibility that mathematical documents allow to (re)define used mathematical notations further corroborates to the complexity of learning mathematics.\markdownRendererInterblockSeparator
{}How would be possible for machines to learn mathematics? One of the major problems is the ambiguity of mathematical expressions (a-c). Natural languages also consist of ambiguous and context-sensitive words. A typical workaround for this problem is to consider the use of lexical databases (e.g.~WordNet~\cite{DBLP:journals/cacm/Miller95}) to identify the most suitable word senses for a given word. However, mathematics lacks of such system, which makes its adoption not feasible at the moment. In~\cite{POM-Tagger} they propose the use of tags, similarly to the POS tags in linguistics, but for tagging mathematical \TeX{} tokens, bringing more information to the tokens considered. As a result, a lexicon containing several meanings for a large set of mathematical symbols is developed. Such lexicons might enable the disambiguation approaches in linguistics to be used in mathematical embeddings in the near future.\markdownRendererInterblockSeparator
{}Furthermore, learning algorithms would benefit from a literature focused in (a) and (d), instead of (b-c). Similar to students who start to learn mathematics, ML algorithms have to consider the structure of the content they learn. It is hard to learn mathematics only considering arXiv documents without prior or complementary knowledge. Usually, these documents represent state-of-the-art findings containing new and unusual notations and lack of extensive explanations (e.g. due to page limitations). In contrast, educational books carefully and extensively explain new concepts. We assume better results can be obtained if ML algorithms are to be trained in multiple stages. A basic model trained on educational literature should capture standard relations between mathematical concepts and descriptors. This model should also be able to capture patterns independently how new or unusual the notations are present in the literature. In 2014, Matsuzaki et al.~\cite{DBLP:conf/aaai/MatsuzakiIAA14} present some promising results to automatically answer mathematical questions from Japanese university entrance exams. While the approach involves many manual adjustments and analysis, the promising results illustrate the different levels of knowledge that is still required for understanding arXiv documents and university entrance level exams. A well-structured digital mathematical library that distinguishes the different levels of progress in articles (e.g.~introductions vs.~state-of-the-art publications) would also benefits mathematical machine learning tasks.\markdownRendererInterblockSeparator
{}Another problem in recent publications, is the lack of standards for properly evaluating MIR algorithms, leading to several publications that present promising results without an extensive evaluation~\cite{DBLP:journals/corr/GaoJYYYT17,DBLP:journals/corr/abs-1803-09123,DBLP:journals/corr/abs-1902-06034}. While ML algorithms in NLP benefit from available extensive training and testing datasets, ongoing discussions about interpretations of mathematical expressions~\cite{SchubotzGSMCG18}, and imprecise standards thwarts research progress in MIR. A common standard for interpreting semantic structures of mathematics would help to overcome the issues of different evaluation techniques. Numerous applications (e.g., search engines) would benefit directly from unified interpretations and representations of mathematical semantics. Therefore, we introduce \markdownRendererEmphasis{Mathematical Objects of Interest} (MOI). Currently, there are three approaches to tokenize and interpret semantics of mathematical expressions: (1) tokenize the mathematical \TeX{} string and tag the tokens with semantic information~\cite{POM-Tagger}, (2) analyze the elements of presentational MathML~\cite{DBLP:conf/sigir/ZanibbiDKT16}, and (3) analyze elements of content MathML~\cite{Schubotz2017,SchubotzGSMCG18}. The Part-of-Math (POM) tagger~\cite{POM-Tagger} proposes a multi-scan approach for mathematical \TeX{} strings that incorporates more semantic information into a parse tree in each iteration. The available first scan creates a parse tree and tags each node with further information about the symbol of each node (similar to POS-tags in linguistics). In~\cite{DBLP:conf/sigir/ZanibbiDKT16}, they generate custom tree representations of presentational MathML (SLT) that allow wildcards for certain positions in the trees and improves search engine results. Applications of (3) are discussed in Section~\ref{sec:related-work}.\markdownRendererInterblockSeparator
{}The goal of MOIs is to combine the advantages of concepts (1-3) and propose a unified solution for interpreting mathematical expressions. We suggest MOIs as a recursive tree structure in which each node is an MOI itself. The current workaround of the problematic example of $\alpha_i$ as an element of the vector $\alpha$ in content MathML is vague and inappropriate for content specific tasks. As an MOI, this expression would contain three nodes, with $\alpha_i$ as the parent node of two leaves $\alpha$ and $i$. While it first seems non-intuitive that $\alpha$, as the vector, is a child node of its own element, this structure is able to incorporate all three components of semantic information of the expression. Hence, an MOI structure should not be misinterpreted as a logical network explaining semantic connections between its elements, but as a highly flexible and lightweight structure for incorporating semantic information of mathematical expressions. \markdownRendererInterblockSeparator
{}We believe that a new standard, such as MOIs, has to be extensively studied and discussed by specialists on the field before its acceptance. Therefore, the introduction of MOIs in this paper is fairly broad and still need to be further investigated. However, we suggest the use of MOIs would assist ML algorithms to simplify and unify their evaluations in MathIR projects.\markdownRendererInterblockSeparator
{}\markdownRendererHeadingTwo{Conclusion and future Work \label{sec:conclusion}}\markdownRendererInterblockSeparator
{}In this paper, we explore how text embedding techniques (e.g. word2vec, PV) are unable to represent mathematical expressions adequately. After experimenting with popular mathematical representations in MIR, we expose fundamental problems that prevent ML algorithms from learning mathematics. We also discover the same problems in several related research projects. Many of these projects show promising examples without extensive evaluations, motivating more researchers to follow the same idea with equivalent fundamental issues.\markdownRendererInterblockSeparator
{}We present concepts for enabling ML algorithms to learn mathematical expressions. Some of these concepts are generally time consuming, such as a lexical database for mathematics. For a concrete contribution we propose the MOIs, a unified solution for interpreting mathematical expressions semantically. We think this is a necessary first step to enable unified evaluations and libraries.\markdownRendererInterblockSeparator
{}In future work, we plan to explore the element distributions of mathematical expressions and compare them with known distributions in linguistics~\cite{Piantadosi2014} for leveraging MOIs'. Preliminary results have shown that distributions of mathematical objects also following Zipf's law, similar to word distributions in natural language. Thus, we are currently analyzing term frequencies and inverse document frequencies, commonly known as tf-idf, of mathematics. This research should help to discover existing meaningful mathematical structures that are already in use in scientific publications. Such structures should be interpreted as MOIs. Therefore, this research will build a base to constructively discuss MOIs.\relax

\paragraph*{Acknowledgments} This work was supported by the German Research Foundation (DFG grant GI-1259-1). We thank Howard Cohl who provided insights and expertise.

\printbibliography[keyword=primary]
\end{document}